%% file: manuscript.tex
\documentclass[fleqn,10pt]{wlscirep}
\usepackage[utf8]{inputenc}
\usepackage[T1]{fontenc}
\usepackage{listings}
\usepackage{xurl}
\usepackage{amsmath}
\usepackage{float}
\usepackage{flafter}
\usepackage[section]{placeins}
\title{Modeling User Redemption Behavior in Complex Digital Incentive Environment: An Empirical Study Using Large-Scale Transactional Data}

\author[1,*]{Akira Matsui}
\author[2]{Takashi Teramoto}
\author[3]{Eiji Motohashi}
\author[4]{Hiroyuki Tsurumi}

\affil[1]{Kobe University, Center for Computational Social Science, Kobe, Japan}
\affil[2]{Chuo University, Faculty of Commerce, Tokyo, Japan}
\affil[3]{Rikkyo University, College of Business, Tokyo, Japan}
\affil[4]{Yokohama National University, Faculty of International Social Sciences, Yokohama, Japan}

\affil[*]{Corresponding author: amatsui@rieb.kobe-u.ac.jp}

\begin{abstract}
  The digital economy implements complex incentive systems to retain users through point redemption.
  Understanding user behavior in such complex incentive structures presents a fundamental challenge, especially in estimating the value of these digital assets against traditional money.
  This study tackles this question by analyzing large-scale, real-world transaction data from a popular personal finance application that captures both monetary spending and point-based transactions.
  We find that point usage is linked to demographics.
  Our analysis using a natural experiment and a causal inference technique reveals that a large point grant stimulated an increase in point spending without a detectable effect on cash expenditure.
  We then find an association between consumers’ shopping styles and their point redemption patterns.
  This study, on a massive real-world economic ecosystem, examines how consumers behave in multi-currency environments, with direct implications for modeling economic behavior and designing digital platforms.
\end{abstract}


\begin{document}

\flushbottom
\maketitle

\makeatletter
\noindent\textbf{Keywords:} \@keywords
\makeatother

\section{Introduction}
\label{sec:intro}

The rise of e-commerce and the proliferation of mobile devices have created a digital economy in which consumers’ searches and purchases are recorded in user logs.
Such detailed data enable researchers and platform administrators to model consumer behavior computationally~\cite{lazer2009computational}.
However, conventional social science methods are often inadequate for analyzing massive consumer behavior recorded through complex incentive designs, and there remains room to extract valuable behavioral insights from large-scale user logs.

Customer management practitioners design incentive structures to induce consumers' purchases. They often employ collaborative ecosystems in which multiple firms jointly participate in value-exchange systems known as loyalty programs (LPs), such as airline alliances or hotel reward alliances.
These digital platforms use platform-specific reward units (e.g., loyalty points) alongside traditional currencies, creating complex multi-party incentive structures. By studying LPs, the literature reveals that they are powerful means to incentivize customers and enhance their loyalty to programs~\cite{blattberg2008database}, but most studies focus on a single platform and they do not study consumers' holistic behavior~\cite{chen2021three}. In recent years, however, partnership loyalty programs (PLPs) have emerged as a dominant form of digital incentive system, in which multiple firms jointly participate and offer or redeem rewards for shared customers~\cite{breugelmans2015advancing}.

Unlike single-firm LPs, PLPs create interconnected networks that allow consumers to earn and redeem rewards across diverse partner firms spanning a wide range of industries~\cite{dorotic2021synergistic}. This multi-partner structure distinguishes PLPs from conventional LPs in a way that is critical for research: because consumers transact across many partner firms, PLP data can reveal a holistic picture of consumer behavior that no single-firm LP dataset can provide. Despite the increasing significance of PLPs, existing research has mainly focused on firm-level sales performance~\cite{bombaij2020when} and cross-purchasing transitions~\cite{dorotic2021synergistic, lemon2009reinforcing}, leaving individual-level redemption behavior largely unexplored with real-world data.

Despite their importance, our understanding of how individual consumers behave in PLPs remains limited, as such studies predominantly rely on laboratory experiments rather than real-world transaction data~\cite{kivetz2006goal, nunes2006endowed, kivetz2002earning}. This gap exists largely because consumer behavior in PLPs is too complex to be studied by traditional methods, particularly regarding unconventional point acquisition and redemption patterns. For example, point redemption plays an important role in their incentive structure, but the monetary value of such ``points'' is not always clear or translatable, which makes comparisons and objective evaluations difficult. More specifically, this study addresses two key gaps: how consumers manage point redemption across multiple industries in PLPs, and how atypical point acquisitions (such as large, one-time grants from campaigns or government initiatives) influence subsequent redemption behavior in real-world settings.
While existing studies examine sign-up or referral bonuses~\cite{kim2021emerging, jin2014when}, large-scale exogenous grants offer a unique opportunity to study stimulus effects.

To overcome the aforementioned challenges, this study leverages detailed consumer behavior histories obtained from a third-party bookkeeping app that includes monetary spending not limited to specific PLPs. This data captures real-world point earning and redemption across several major PLPs. The PLP market in Japan is particularly dynamic and expansive, characterized by intense competition~\cite{MizuhoBank2025PointEconomy, steinberg2025incentives}. It has also grown substantially, as the annual value of loyalty points issued by major Japanese companies already exceeds JPY 1 trillion and is projected to rise to over JPY 1.2 trillion by FY2025~\cite{Tomita2023LoyaltyPoints}.
Furthermore, PLP platforms in Japan serve as channels for government policy initiatives, distributing rewards for actions like digital ID registration or adopting contactless delivery \cite{fujiki2025cashless, sekine2022going}, making PLPs central hubs for reward exchange among firms, institutions, and consumers. We model and investigate the data by conducting natural experiments arising from government policies and by classifying shopping styles using embedding representations learned from vast purchasing histories.

By investigating the interplay among consumer attributes, point acquisition mechanisms, shopping habits, and redemption choices, we aim to provide novel insights into consumer behavior within this complex ecosystem. Theoretically, this research extends the understanding of LPs to the collaborative context and provides empirical evidence on consumer rationality (or irrationality) in point usage, particularly concerning mental accounting and responses to exogenous stimuli, thereby advancing marketing theory. Managerially, our results offer practical implications for optimizing PLP design, targeting promotions, and evaluating the effectiveness of point-based public policies.

We address the following three research questions (RQs):
\begin{itemize}
  \item \textbf{RQ1} {\it Do demographic attributes explain point redemption patterns?}
  \item \textbf{RQ2} {\it Does receiving a significant point grant stimulate point redemption?}
  \item \textbf{RQ3} {\it Is the diversity of consumers' day-to-day store visits (exploration vs.\ exploitation) associated with how they allocate point redemptions across PLP categories?}
\end{itemize}

By answering these questions, this study provides empirical analysis on consumer engagement with PLPs to generate insights for both academic research and practical management of PLPs. The RQ1 seeks to understand how point redemption for specific goods or services varies across customer segments, which can inform targeted marketing and program design. In turn, we answer RQ2 to examine the impact of non-purchase-related point acquisitions, such as government incentives. This examination helps evaluate the effectiveness of point programs as behavioral triggers and can inform optimal incentive design. Investigating RQ3 involves comparing consumers who explore diverse retailers versus those who exploit a few familiar ones. This comparison can shed light on cross-selling potential and loyalty mechanisms within PLPs.
To make the rest of the paper easier to read, we introduce the key terms used in this study in Table~\ref{tab:plp_terms}.

\input{table1.tex}

\section{Related Research}
\label{sec:related}

Our study builds upon and extends two primary streams of literature: partnership loyalty programs (PLPs) and loyalty programs (LPs) in general. This review contextualizes our research by highlighting existing knowledge and the specific gaps our study addresses.

\subsection{Consumer Behavior and Firm Performance in PLPs}
\label{subsec:related_plp}

Research on PLPs has explored their impact from both firm and consumer perspectives, yielding a mixed set of findings. From the firm's standpoint, some studies suggest that retailers participating in PLPs might experience lower sales compared to those operating sole-proprietary programs~\cite{bombaij2020when}. This outcome could be attributed to factors such as diluted brand association or intensified competition among partners. From the consumer perspective, PLPs have demonstrated an ability to achieve higher customer penetration rates than firms not affiliated with PLPs~\cite{sharp1997loyalty}. PLPs can also induce cross-purchasing behavior, particularly among customers who frequently shop at specific partner retailers~\cite{lemon2009reinforcing}. Furthermore, synergistic effects have been observed: joint promotions by multiple partners can elicit stronger customer responses than single-firm promotions~\cite{dorotic2011do}; PLP mobile application usage can increase spending~\cite{kim2015effects}; and shifting rewards towards instant monetary value can boost redemption activity~\cite{wang2018when}.

However, the literature also includes studies reporting less positive or neutral effects. For instance, some research finds no significant difference in repeat purchase behavior between partnering and non-partnering retailers \cite{sharp1997loyalty}, or no discernible difference in purchase volume between LP users and non-users within PLPs \cite{villacemolinero2016multi}. In addition, cross-purchasing may not materialize if partner offerings lack complementarity \cite{lemon2009reinforcing}, and in some cases, cannibalization among partners can outweigh synergistic benefits \cite{dorotic2021synergistic}. These varied findings underscore the complexity of PLP dynamics and highlight the need for deeper insights into consumer engagement that extend beyond simple purchase metrics, a gap this study aims to address by examining detailed redemption behaviors.

\subsection{Reward Earning and Redemption in LPs}
\label{subsec:related_earning_redeeming}
The design of rewards is a central element in determining LP effectiveness \cite{blattberg2008database}. Research in this area distinguishes among various reward characteristics that influence earning behavior. Rewards can be direct (e.g., cashback at the same retailer) or indirect (e.g., discounts at unrelated partners), with customers generally showing a preference for direct rewards \cite{keh2006do, yi2003effects}. Rewards can also be immediate (e.g., checkout discounts) or delayed (e.g., points accumulating for future redemption); immediate rewards are often preferred due to time discounting effects \cite{yao2012determining, keh2006do}.

Beyond these rational preferences, behavioral economics research, frequently employing laboratory experiments, has illuminated non-rational aspects of reward earning. The first notable concept in this line of literature is ``mental accounting,'' developed by~\cite{thaler1985mental}, which highlights consumers' deviations from rational decision-making, influenced by mental shortcuts such as cognitive biases. Other notable examples include the ``goal gradient effect,'' where effort increases as a reward threshold approaches \cite{kivetz2006goal}; the ``endowed progress effect,'' where initial bonus points boost motivation \cite{nunes2006endowed}; and the ``idiosyncratic fit heuristic,'' where perceived value increases if consumers feel advantaged \cite{kivetz2003idiosyncratic}.

Research on reward redemption similarly explores a variety of influencing factors. Higher redemption thresholds tend to decrease redemption rates \cite{dreze1998exploiting}, whereas higher monetary value rewards or lower thresholds can encourage redemption \cite{dreze2004using, dreze2011recurring}. Some studies using real-world data indicate that consumers may hoard points for cognitive or psychological reasons \cite{stourm2015stockpiling}. The type of reward (social versus economic) might not significantly affect redemption intentions \cite{noble2014accumulation}, while the ease of calculating a reward's value can enhance such intentions \cite{wang2018when}. Redemption choices can also be influenced by the perceived effort involved in earning the reward; high-effort rewards might be spent on luxuries, while low-effort rewards are often directed towards necessities \cite{kivetz2002earning}. Furthermore, rewards detached from personal effort, such as gifts or lottery winnings, might lead to weaker redemption choices \cite{kivetz2005promotion}. Our study extends this line of inquiry by examining large-scale, exogenous point grants in a real-world PLP context.

Existing research on redemption behavior, especially concerning behavioral biases or responses to non-standard point acquisition methods (such as large exogenous grants), predominantly relies on laboratory experiments. Consequently, studies utilizing real-world transaction data to examine redemption behavior following such significant, externally driven events are scarce~\cite{stourm2015stockpiling}.
Our research aims to fill these gaps by analyzing actual PLP transaction data. It addresses both the limited understanding of cross-partner redemption patterns in PLPs and the lack of real-world evidence on responses to non-standard point acquisitions. 
We specifically focus on redemption patterns across various partners and the impact of a large-scale government point grant, thereby connecting point utilization to broader consumption habits and contributing empirical evidence from a real-world setting.

\section{Data and Methods}
\label{sec:methods}

We empirically analyze consumer behavior within Japan's PLP market using transaction records from a widely used personal finance application. This section details the data source and the analytical methods employed to address our research questions.

\subsection{Data}
\label{subsec:data}

The primary data source for this study is the Zaim application, a popular smartphone-based bookkeeping tool in Japan. Users can manually input their daily expenditures and income or automatically sync data from bank accounts, credit cards, and various loyalty programs via API connections. Crucially for this research, Zaim allows users to track both the ``income'' (earning) and ``redeem'' (spending) of points associated with numerous major Japanese PLPs alongside regular cash or card transactions. This provides a unique, integrated view of consumers' financial activities, including their engagement with multiple PLPs. As Zaim is an independent third-party application with no capital ties to any specific PLP operator, the data is less likely to be biased towards a particular program compared to data sourced directly from an LP provider. The self-recording nature for budgeting purposes suggests a reasonable level of data reliability.

For data preprocessing, we conducted several steps to ensure data quality and consistency. First, we removed transactions with missing essential information, such as store names or transaction amounts. Next, we standardized store names and category labels across different merchants and excluded extreme outliers in transaction amounts, defined as transactions falling outside the 1st and 99th percentile thresholds. From a privacy perspective, all user data were anonymized prior to analysis, and researchers had no access to personally identifiable information. The final analysis sample consists of 41,934 customers with purchase transactions and 14,026 customers with point transactions. The dataset analyzed includes over 5.5 million unique purchase transactions and 5.9 million point transactions. The distributions of transaction amounts and frequencies across demographic groups are shown in Fig.~\ref{fig:rq1_basic_stats} of the Supplementary Information (Section~\ref{app:rq1_basic_stats}).

\subsection{Methods}
\label{subsec:methods_analysis}
Drawing on detailed transaction histories from a personal finance application, this paper presents three empirical analyses, each tailored to one research question. First, for RQ1, we model how demographic attributes are associated with category-level point redemption patterns using topic-based representations and multivariate regressions. Second, for RQ2, we exploit a quasi-experimental government point grant and estimate its causal effect on subsequent point redemption and monetary spending using a doubly robust Difference-in-Differences design. Third, for RQ3, we quantify users' exploration--exploitation tendencies from store-visit trajectories and examine how these tendencies are associated with the allocation of point redemptions across PLP categories.

\subsubsection{Labeling point transaction records}\label{subsec:method_rq1}
To investigate how point redemption patterns relate to customer attributes (RQ1), we first process and analyze the textual descriptions associated with point redemption transactions. This involves preprocessing the text data by removing common noise words (e.g., ``receipt,'' ``purchase'', or ``sell'') and extremely frequent terms. We then apply Latent Dirichlet Allocation (LDA)~\cite{Blei2003} to identify latent topics or contexts characterizing point usage. To select the number of topics, we prioritized semantic coherence and practical interpretability over statistical heuristics like the elbow method. We ran LDA models with 30, 50, and 100 topics. For each model, we extracted the top 100 words for each topic based on their conditional probability. We then generated labels for each topic by providing these word sets to a large language model (GPT-4o), along with a minimal description of the context (i.e., that they were results from a topic model of point transaction records). 
Rather than adopting these machine-generated labels directly, we manually reviewed and validated each candidate label to confirm that it was consistent with the topic's top-ranked words.
Through this manual review process on the generated labels, we concluded that the model with 30 topics provided the most intuitively coherent and practical fit. 
We then construct individual-level feature vectors that contain their spending patterns based on the identified topics, which represent customers' redemption or spending transactions. We then cluster these user profile vectors using the k-medoids clustering algorithm to identify distinct groups of users with similar point redemption patterns, setting the number of clusters to 5 as determined by the elbow method. We examine these representations of customer behavior in Section~\ref{sec:results_rq1}.

\subsubsection{Leveraging natural experiment to understand the effect of point redemption by large grants}\label{subsec:method_rq2}

To assess whether earning points stimulates subsequent spending (RQ2), we leverage a natural experiment provided by a Japanese government's policy. Specifically, we utilize data from the ``second My Number Point campaign'' ({\it My Number Point Dai-2 Dan}) in late 2022~\cite{digitalAgency2022MynaPoints}. 
This was a government-led nationwide campaign conducted to incentivize the adoption of a specific administrative service by providing reward points. During this campaign, eligible individuals could receive a one-time grant of up to JPY 7,500 in points from their chosen PLP upon completing certain actions (e.g., linking their health insurance card). This campaign was part of Japan's broader digital transformation initiative, aimed at encouraging citizens to adopt digital administrative services and promote cashless payments. The points could be used at any participating PLP partner, making it a unique opportunity to study how consumers respond to large, exogenous point grants in a real-world setting.

To study this question, we employ a Difference-in-Differences (DID) design. Our data show that a subset of users received a one-time grant of JPY 7,500 points through this Japanese government campaign. Specifically, we identify users who received this grant on September 23, 2022 (Figure~\ref{fig:did_treatment_group}). We define these users as the treatment group ($N=349$ in our sample), who received the same JPY 7,500-point grant within a narrow time window. The ``control group'' comprises users ($N=12,192$) who did not receive this specific grant during the same period. We compare the change in point redemption behavior (e.g., total points redeemed per month/week) before and after the intervention window for the treatment group to the corresponding change in the control group. While a natural experiment mitigates some concerns, it does not eliminate the risk of selection bias. For example, users who opted into the campaign may systematically differ from those who did not, potentially violating the parallel trends assumption essential for standard Difference-in-Differences (DID) analysis. To address this potential issue, we employ the doubly robust estimation method proposed by \cite{SantAnnaZhao2020}. This approach yields consistent estimates for the treatment effect if either a propensity score model (for treatment assignment) or an outcome regression model is correctly specified, without requiring both. We estimate propensity scores based on pre-intervention observable characteristics, including demographics (age, family composition, region) and behavioral variables (e.g., total cash spending levels, prior point usage frequency, diversity of spending). This approach enhances the robustness of our causal estimate of the average treatment effect on the treated (ATT) of the point grant on subsequent point redemption.

\subsubsection{Quantifying Customers' Exploration and Exploitation and its connection to point redemption}\label{subsec:method_rq3}
To explore the relationship between consumers' general \textit{shopping style} and their point transactions (RQ3), we first quantify each user's tendency towards exploration and exploitation in their overall purchasing behavior and then examine its connection to point redemption. To capture their shopping style, we employ the exploration-exploitation framework~\cite{Kim2024PNAS,GOMEZZARA2024108014}. In this concept, \emph{explore} refers to the tendency of consumers to try new or diverse stores/merchants compared to their past behavior, while \emph{exploit} refers to the tendency to repeatedly visit familiar or a narrow set of stores/merchants. This framework helps explain the breadth of their point redemption within the PLP network.

Following methodologies used in analyzing consumption variety (e.g., \cite{Kim2024PNAS, mok2022dynamics}) and drawing on embedding techniques \cite{Kim2024PNAS,waller2021quantifying}, we learn vector representations (embeddings) for each store/merchant present in the transaction data using a Skip-gram with Negative Sampling (SGNS) model, similar to word2vec \cite{mikolov2013efficient, mikolov2013dist,levy2014dependency}, trained on sequences of stores visited by users. The analysis proceeds in three steps.

\textbf{Learning vector representations of stores: }
Following \cite{waller2021quantifying, Kim2024PNAS} we treat the ordered list of stores each user visits as an analogue of a sentence and learn dense embeddings with a Skip-gram with Negative Sampling (SGNS) model \cite{mikolov2013efficient}. Concretely, we construct for every user $i$ the sequence $\langle s_{i1}, s_{i2},\dots\rangle$ of stores visited in chronological order (multiple visits to the same store on the same day are retained). All sequences are concatenated into a single corpus on which we train word2vec with embedding dimension $d=128$, context window $=5$. The model yields a vector $\mathbf{e}_{s}\in\mathbb{R}^{d}$ for each store $s$.

To construct customer-level representation, by following~\cite{Kim2024PNAS}, for user $i$ on day $t$, let $S_{it}$ be the multiset of stores recorded in the transaction log. The average store vector
\[
  \mathbf{u}_{it}=\frac{1}{|S_{it}|}\sum_{s\in S_{it}}\mathbf{e}_{s}
\]
summarizes that day's shopping pattern. To capture short-run routine we compute the rolling mean over the previous six days,
\(
  \bar{\mathbf{u}}^{(6)}_{i,t-1}=6^{-1}\sum_{\tau = t-6}^{t-1}\mathbf{u}_{i\tau}.
\)
A higher cosine similarity between $\mathbf{u}_{it}$ and $\bar{\mathbf{u}}^{(6)}_{i,t-1}$ indicates exploitation; its complement indicates exploration \cite{Kim2024PNAS}. Formally,
\[
  \mathrm{Exploit}_{it}=\cos(\mathbf{u}_{it},\bar{\mathbf{u}}^{(6)}_{i,t-1}),\quad
  \mathrm{Explore}_{it}=1-\mathrm{Exploit}_{it}.
\]

\section{Results}
\label{sec:results}

This section presents the empirical findings corresponding to our three research questions. We first examine the relationship between point redemption patterns and customer attributes (RQ1), then assess the impact of a large point grant on subsequent spending (RQ2), and finally investigate the correlation between store usage habits and point redemption diversity (RQ3).

\subsection{Point Redemption Patterns and Customer Attributes (RQ1)}
\label{sec:results_rq1}

\begin{figure}[!htbp]
  \centering
  \includegraphics[width=0.75\linewidth]{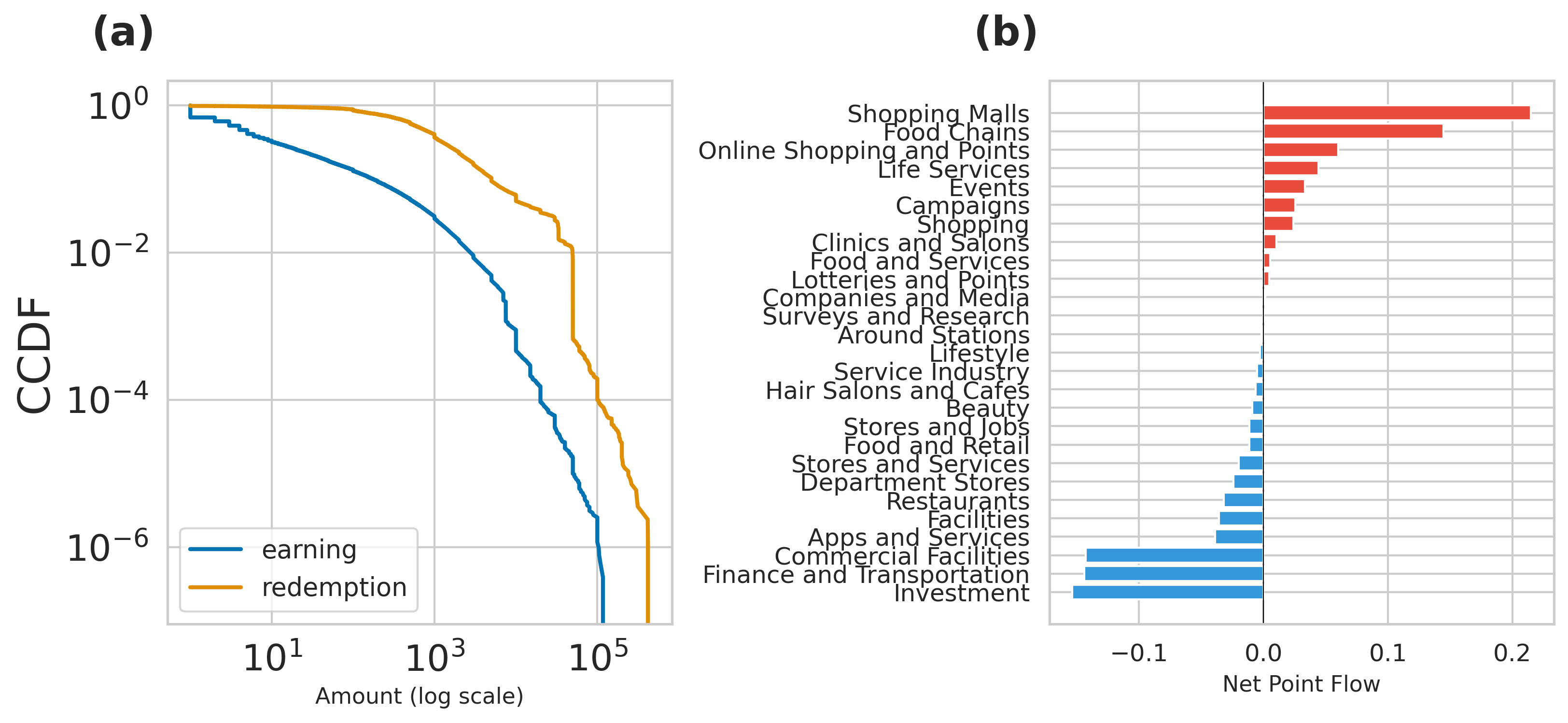}
  \caption{Overview of point transaction patterns. (a)~Complementary cumulative distribution (CCDF) of point earning and redemption amounts, showing that redemptions exhibit a heavier tail. (b)~Net point flow across categories, computed as the difference between the user-level normalized earning share and the normalized redemption share for each topic, averaged across all users. A negative value indicates that customers redeemed a larger share of their points than they earned in that category; a positive value indicates the opposite.}
  \label{fig:rq1_basic_overview}
\end{figure}

Our initial analysis characterizes customer spending behavior across both monetary and point-based transactions to understand how points are actually used within the PLP ecosystem. More specifically, we describe how customers in PLPs manage their spending and points in their purchases, and then we investigate their spending behavior with respect to their demographics. We first observed a fundamental asymmetry between how points are earned and spent. While points are typically earned in small, incremental amounts from various purchases, they are often redeemed in larger, consolidated sums. As shown in Figure~\ref{fig:rq1_basic_overview}(a), point redemption exhibits a heavier tail than point earning, indicating that while most redemptions are for smaller amounts, high-value redemptions are rare. This structure incentivizes customers to remain engaged within a single PLP to accumulate enough points for meaningful redemptions.

Our analysis of net point flows (earnings minus redemptions) by category reveals that transactions are not uniform; instead, some categories serve primarily as sources of point earnings, while others are destinations for point redemption (Figure~\ref{fig:rq1_basic_overview}(b)). This analysis also provides a key managerial insight: a distinct group of users directs their point redemption towards financial services. This is a notable trend in the Japanese market, where major PLPs often own brokerage firms, allowing customers to invest their points in financial products like mutual funds or stocks. A prime example is the Rakuten ecosystem, where customers earn points across a vast network of services and are then encouraged to redeem them within the same network, including for investments.

In these PLPs, customers can apply as little as one point to a purchase by combining it with cash. Yet most do not use points at the point of sale and instead accumulate them for later, larger redemptions, even though hoarding yields no interest or added benefit. 
The same asymmetry appears across categories, with customers redeeming points for financial services (e.g., mutual-fund purchases) rather than where they were earned, such as restaurants.
This suggests that customers do not treat points as equivalent to cash, consistent with mental accounting~\cite{thaler1985mental}.

\begin{figure}[!htbp]
  \centering
  \includegraphics[width=0.6\linewidth]{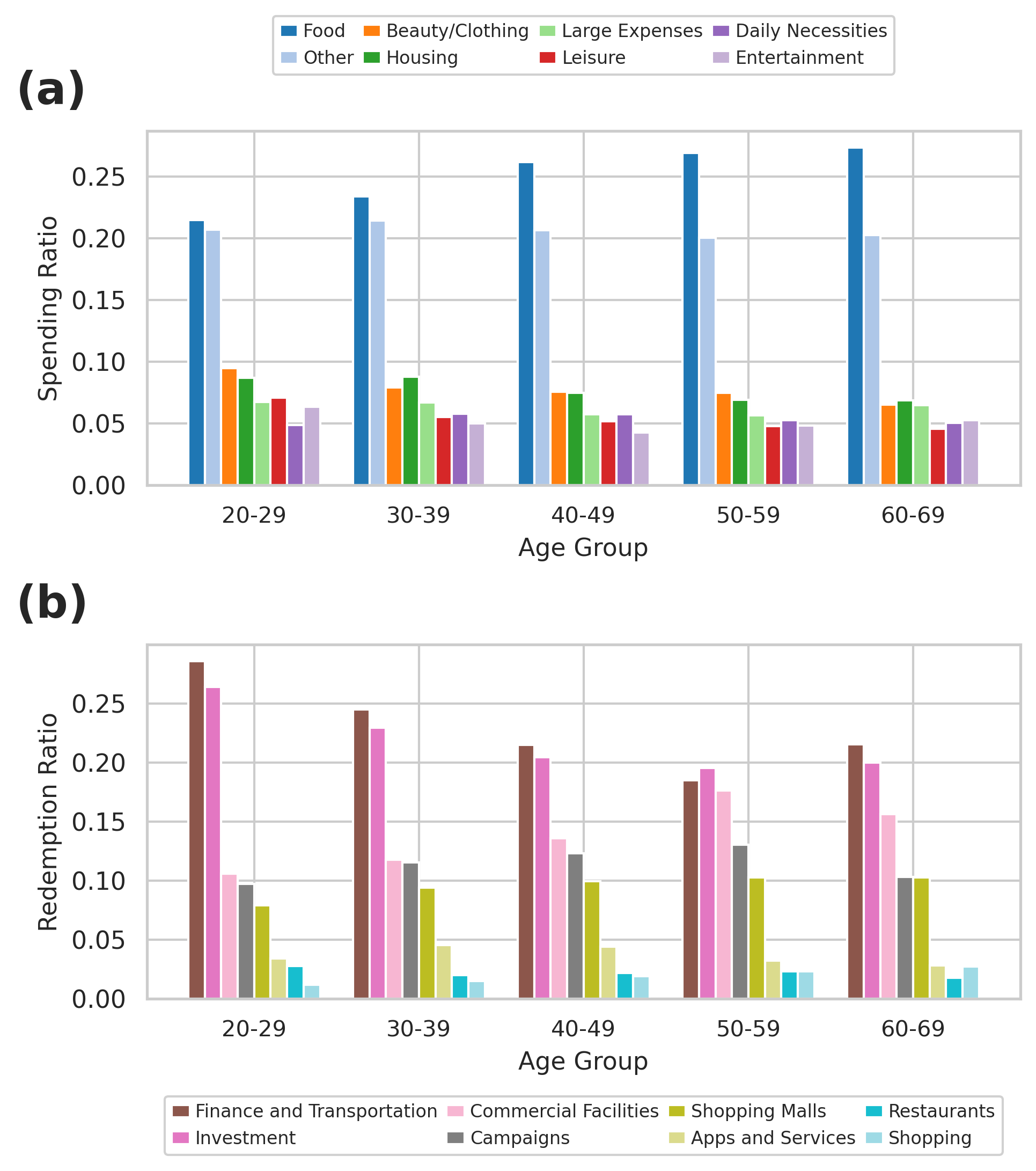}
  \caption{Distribution of spending across categories by age group. Bar charts show the average (a)~spending share and (b)~redemption share, where each bar represents the mean share for an age group.}
  \label{fig:rq1_spe_red_share}
\end{figure}

Next, we study how customer spending behavior relates to demographics, a critical factor for both marketing and economic policy. As shown in Figure~\ref{fig:rq1_spe_red_share}, we analyzed the distribution of monetary and point redemption across categories by age group.

The results indicate that spending patterns differ across age cohorts, although the differences are not always large. Monetary expenditures vary only modestly. For example, customers in their 20s spend slightly more on ``Beauty/Clothing,'' ``Leisure,'' and ``Entertainment.'' In contrast, point-redemption behavior shows clearer differences across age groups (Figure~\ref{fig:rq1_spe_red_share}). We also find differences in the redemption categories of ``Finance and Transportation'' and ``Investment'' (Figure~\ref{fig:rq1_spe_red_share}(b)). The share of points spent in each category differs by approximately 6 to 7 percentage points between customers in their 60s and those in their 20s. Younger customers allocate a much larger share of their points to these areas. This suggests that redemption priorities shift from financial services toward discretionary goods as consumers age.

\begin{figure}[!htbp]
  \centering
  \includegraphics[width=0.99\linewidth]{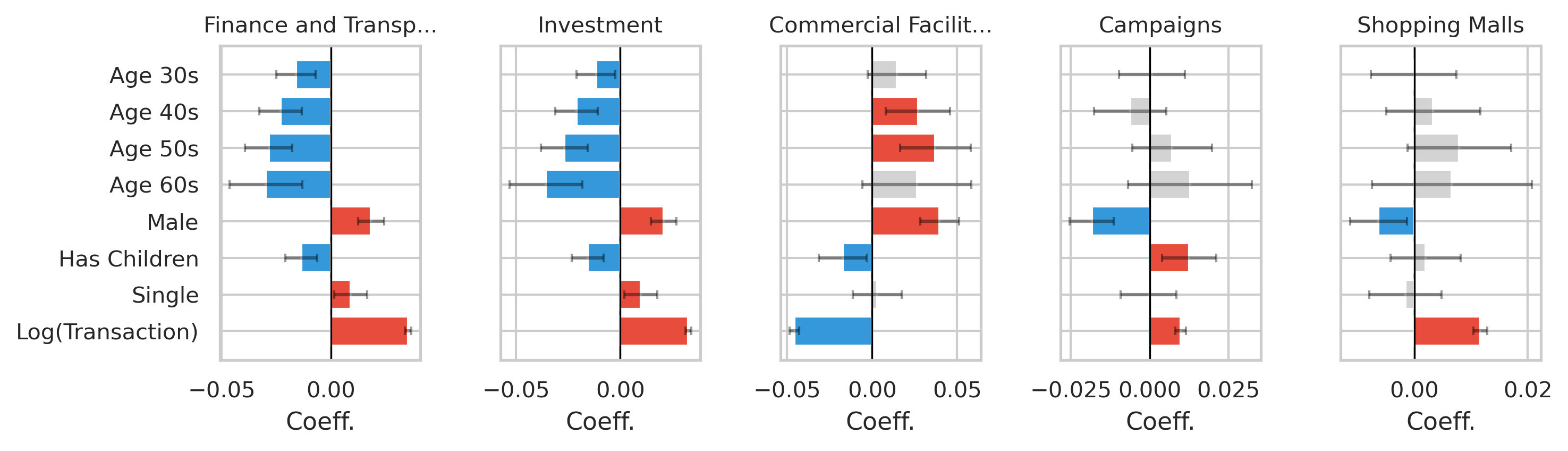}
  \caption{OLS regression coefficients for demographic predictors of point redemption share across the five largest redemption categories. The reference category for age is users in their 20s. Error bars indicate 95\% confidence intervals. The effect of demographics is category-dependent: age and gender dominate in financial categories, while family composition matters more for campaigns.}
  \label{fig:rq1_regression}
\end{figure}

To study this association between point transaction and the demographic factors, we estimate OLS regression models in which the dependent variable is each user's redemption share in the five largest categories, and the covariates include age group, gender, family composition, and total transaction volume. The estimates show that demographic patterns differ by category. In ``Finance and Transportation'' and ``Investment,'' younger users tend to allocate a larger share of points, while in ``Commercial Facilities'' the middle-aged group shows the highest shares. In ``Campaigns,'' age shows little association, whereas gender and family composition are more salient. Overall, demographic attributes are associated with point redemption, but the strength and direction of these associations vary across categories (Figure~\ref{fig:rq1_regression}).

These findings suggest that while PLP point redemption is widespread across demographics, the specific ways points are utilized are not random but align predictably with user life stages and associated needs, offering potential avenues for targeted promotions within PLPs.

\FloatBarrier

\subsection{Effect of Exogenous Point Grant on Subsequent Spending (RQ2)}
\label{sec:results_rq2}

Our findings for RQ1 revealed that monetary spending and point redemption share similar statistical patterns. However, this similarity does not guarantee that customers treat points and money identically. According to the theory of mental accounting~\cite{thaler1985mental}, even if points and cash have the same purchasing power, their perceived value can differ. For instance, if consumers do not perceive points as equivalent to cash, they might be more inclined to spend them immediately rather than save~\cite{ChenPointRedemption2024,stourm2015stockpiling}. This hypothesis leads us to ask whether customers behave differently when they receive points compared to when they receive cash. Because the 7,500-point grant examined here was distributed by the Japanese government as part of a public policy initiative, RQ2 addresses a question analogous to a canonical problem in fiscal policy: {\it What fraction of an exogenous transfer is spent, and what fraction is saved?} In macroeconomics, this ratio is known as the marginal propensity to consume. By ``stimulate,'' we therefore mean whether the grant causes incremental point spending relative to a counterfactual in which the grant was not received, not whether it induces spending in excess of the granted amount.

This question has two important implications. First, for administrators of PLPs, it is important to know how many of the points given through marketing campaigns are actually used within their own program. Second, for policymakers, it is important to understand how much of the given points people spend and how much they ``save''. This helps measure how effective a policy is by looking at how much it changes people's spending.

To answer this, we investigate the effect of a 7,500-point (equivalent to JPY 7,500) grant from a Japanese government initiative\footnote{The second My Number Point campaign, as described in Section~\ref{sec:methods}}. We employ a difference-in-differences (DID) methodology with doubly robust estimation to mitigate selection bias, comparing the change in spending for customers who received the grant against those who did not. Our primary outcome of interest is the total value of points redeemed in the month following the grant distribution.

\begin{figure}[!htbp]
  \centering
  \includegraphics[width=0.95\linewidth]{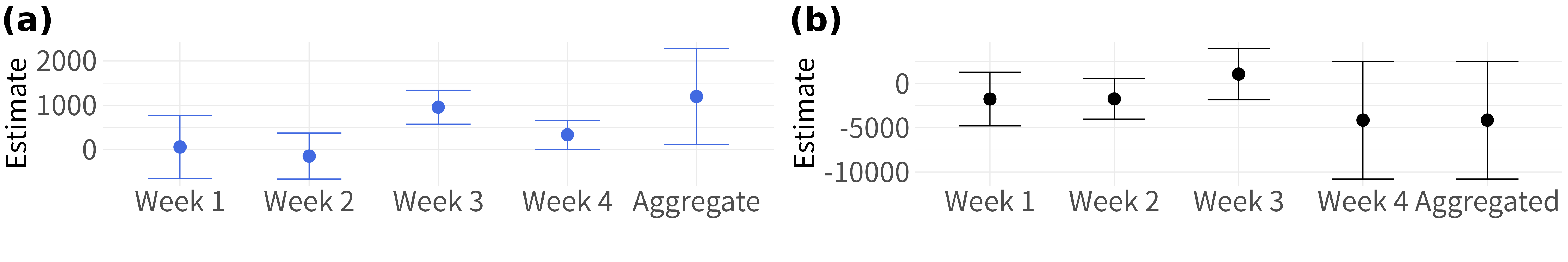}
  \caption{Difference-in-Differences estimation of the effect of the JPY 7,500 point grant. (a)~Effect on monthly point redemption. (b)~Effect on monthly monetary expenditure. Both panels show the estimated ATT over time relative to the intervention period (Week~0). Error bars indicate 95\% confidence intervals. The ``Aggregated'' result represents the total effect across the entire post-intervention period.}
  \label{fig:did_combined}
\end{figure}

The results indicate that the point grant significantly stimulated point redemption. In the first month post-intervention, users in the treatment group redeemed, on average, 1,203.15 (JPY equivalent) more in points than those in the control group ($p = 0.0302$). This represents approximately 16\% of the total JPY 7,500 grant being spent within the first month. As depicted in Figure~\ref{fig:did_combined}(a), the average treatment effect shows a clear spike following the intervention. A weekly analysis reveals a slight delay, with redemption peaking in the third week (an increase of 955.04 points, $p < 0.001$).

This spending share can be interpreted as the marginal propensity to consume out of a point windfall. As with cash transfers, recipients appear to spend only part of the grant at first and keep most of it for later use, even though saved points yield no financial return. Several mechanisms may account for this conservative pattern: point stockpiling for a future large redemption~\cite{stourm2015stockpiling, ChenPointRedemption2024}, mental accounting that categorizes government-granted points differently from self-earned points~\cite{thaler1985mental, kivetz2005promotion}, and a cash-like savings motive in which consumers treat points as a store of value despite the absence of financial returns. The weekly analysis further supports this interpretation: redemption peaked in the third week ($+955.04$ points; $p<0.001$), suggesting a deliberation period before planned spending rather than impulsive consumption.

On the other hand, our analysis finds no corresponding effect on monetary spending. The DID analysis reveals no statistically significant difference in cash expenditures between the treatment and control groups following the grant (Figure~\ref{fig:did_combined}(b)). This null result suggests that the awarded points did not directly substitute for cash. Consequently, while point grants can stimulate point-based consumption, they may not be an effective policy tool for inducing broader monetary spending.

The DID analysis suggests that the grant led to additional point redemption, but since not all of the 7,500 points (JPY equivalent) were redeemed immediately, a natural follow-up question arises: {\it How long did it take for recipients to redeem the full grant amount?} To answer this, we compare the time taken by the treatment group to spend a cumulative 7,500 points (JPY equivalent) post-intervention against a matched control group. This control group was constructed using propensity score matching based on demographic characteristics and prior spending patterns to ensure a valid comparison. As shown in Figure~\ref{fig:ps_matching_spending}, the treatment group exhausted the 7,500-point value more quickly than the control group, indicating a sustained, faster spending pattern after receiving the grant.

\subsection{Store Usage Patterns and Point Redemption Diversity (RQ3)}
\label{sec:results_rq3}

Our findings for RQ2 revealed no direct substitution effect between the point grant and monetary spending. However, as RQ1 suggested a link between demographic traits and point usage, our final research question (RQ3) investigates another dimension of consumer heterogeneity: shopping style. 
Specifically, we examine whether customers’ general shopping behavior is associated with their point-redemption patterns. We define shopping behavior as either exploratory or exploitative.

To do so, we quantify each customer's shopping behavior in terms of exploration and exploitation (E/E) using the embedding-based method described in Section~\ref{subsec:method_rq3}. In parallel, we characterize their point redemption patterns by clustering users based on their spending distribution across the LDA topics identified previously. The interpretation of these clusters is presented in Table~\ref{tab:point_cluster_labels}.

\begin{figure}[!htbp]
  \centering
  \includegraphics[width=0.8\linewidth]{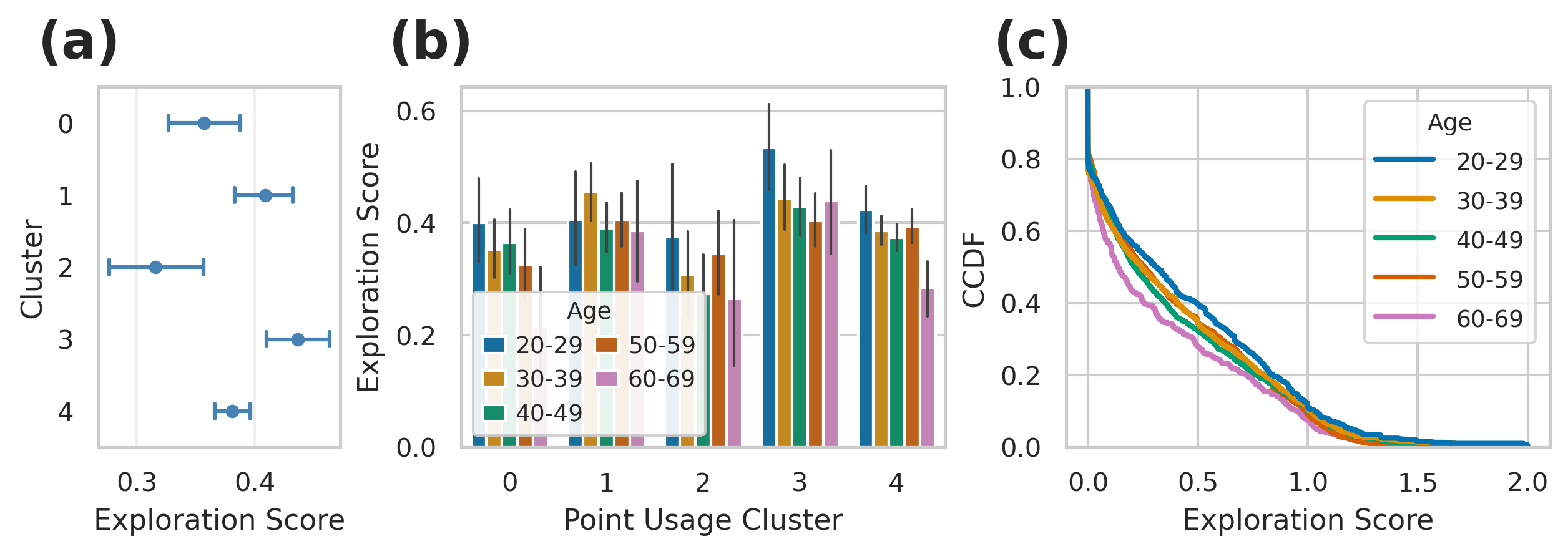}
  \caption{Exploration--exploitation analysis of shopping behavior. (a)~Average exploration scores across different point usage clusters; error bars represent 95\% confidence intervals. (b)~Average exploration score by point usage cluster and age group, showing that exploration tendencies vary more with age than with point usage category. (c)~CCDF of exploration scores by user age group, showing that the young cohort (20s) tends to be more exploratory.}
  \label{fig:rq3_exploration}
\end{figure}

\input{table2.tex}

Our analysis reveals a clear relationship between a user's shopping style and their point usage patterns. As shown in Figure~\ref{fig:rq3_exploration}(a), customers with more targeted point usage (e.g., food-chain and beauty-service users in Cluster 2) exhibit lower average exploration scores. This suggests that exploitative shoppers, who frequent a narrow set of familiar stores, also concentrate their point redemptions, while exploratory shoppers redeem points more broadly.

We also find that these exploration tendencies are strongly associated with demographic attributes, particularly age. Figure~\ref{fig:rq3_exploration}(b) illustrates that exploration scores differ significantly across age groups, even within the same point usage cluster. In fact, age appears to be a more pronounced predictor of a user's exploration score than their point redemption category (Figure~\ref{fig:rq3_exploration}(c)). The young cohort (20s) consistently shows higher exploration scores, indicating a greater propensity to try new stores.
Pairwise Welch's $t$-tests with Bonferroni correction showed that the 20--29 age group scored significantly higher than the 60--69 group (mean difference $=0.108$, $p<0.001$). The 30--39 and 50--59 groups also scored significantly higher than the 60--69 group ($p<0.05$).

As a robustness check, we also measured store diversity using the normalized Shannon entropy of store visits (SI~\ref{app:entropy}), which yielded consistent results. Unlike the embedding-based exploration measure used in the main analysis, entropy treats all stores as independent categories and does not account for similarity between stores (e.g., two supermarkets are considered as different as a supermarket and a travel agency). Pairwise Welch’s t-tests show that younger cohorts generally exhibit higher entropy than older cohorts, indicating that younger users distribute transactions across a larger number of distinct merchants, while older users concentrate spending in a smaller set of frequently visited stores. However, because entropy ignores semantic similarity between stores, it primarily captures dispersion across store identifiers rather than qualitative diversification of consumption. Therefore, this entropy-based analysis should be interpreted as a complementary measure to the embedding-based exploration metric reported in the main text. These results suggest that when diversity is measured using a simple entropy-based approach that does not account for qualitative similarity between stores, younger users appear to exhibit more diverse purchasing patterns. 

To assess the sensitivity of these results to the embedding specification, we re-trained the SGNS model at three embedding dimensions ($d=64$, $128$, and $300$) and recomputed exploration scores from scratch. The Spearman rank correlations between all pairs of dimension settings exceeded $\rho=0.99$, confirming that our findings are robust to this modeling choice (see Supplementary Material).

These findings confirm a meaningful link between a consumer's general shopping style and their point redemption behavior within a PLP. Exploratory consumers leverage the breadth of the PLP network, while exploitative consumers use points primarily within their established routines. This has implications for cross-selling strategies and understanding loyalty dynamics within collaborative programs.

\FloatBarrier

\section{Discussion and Conclusion}\label{sec:d_c}

This study provided empirical insights into consumer behavior on PLPs using large-scale, real-world transaction data, addressing three key research questions.
First, we found that points obtained from PLPs exhibit similar distributional patterns to monetary spending across categories,
reflecting the broad redemption opportunities within PLPs. At the same time, however, recipients of an exogenous point grant
redeemed only a fraction of the awarded amount in the short term, exhibiting a savings-like motive analogous to what is observed with cash transfers.
We also found that some demographic attributes are associated with redemption patterns, and identified groups of customers who adopt specific redemption strategies. We find a group in which customers allocate a certain portion of points to financial services. Other groups exhibit a wide variety of behaviors, suggesting a differential perception of point utility across segments. Second, we revealed that a certain amount of exogenous point grants can stimulate subsequent point redemption. On the other hand, customers, on average, do not spend all the granted points; rather, they tend to save them. This finding aligns with the standard discussion of marginal propensity to consume in economic policy and suggests that providing points does not always lead to irrational spending behavior. 
Notably, the approximately 16\% redemption rate in the first month is consistent with the conservative spending patterns observed in cash-based fiscal stimulus programs. 
This suggests that consumers apply a savings motive to points just as they do to money, even though hoarded points do not generate interest or other financial returns.
Third, we found an association between consumers' general shopping behavior (exploration vs. exploitation) and their point redemption behavior within the PLP. This applies the exploration-exploitation framework to point redemption and contributes to the debate on synergy versus cannibalization in PLPs. 
The finding that age is a stronger predictor of exploration than point-usage cluster warrants further discussion.
Our results suggest that life-stage factors (e.g., age) likely drive this pattern: established households tend to develop stable multi-store routines, whereas younger consumers in transitional life stages exhibit broader and more exploratory shopping behavior.

Regarding theoretical contributions, our study contributes to the LP literature by providing detailed empirical evidence on redemption behavior within the understudied PLP context using real-world data. We offer empirical support for mental accounting in point usage and demonstrate the behavioral impact of large exogenous point grants outside a laboratory setting. Linking exploration-exploitation tendencies to point redemption patterns offers a new lens for understanding consumer engagement with multi-partner programs.

In terms of managerial and policy implications for PLP operators and partner firms, our findings underscore the importance of understanding redemption flows across the network of PLPs. Segmenting users by demographics (RQ1) and shopping style (RQ3) can inform targeted promotions and partner recommendations. Exploitative users might be targets for deepening engagement with preferred partners, while exploratory users represent opportunities for cross-partner discovery. For policymakers, the effectiveness of point-based incentives (RQ2) is supported, but with an important caveat: point grants stimulate consumption within the point economy but do not appear to substitute for cash spending.
This means that point-based policies may be effective for promoting digital adoption or directing consumers toward specific services. However, their usefulness as a general fiscal stimulus that aims to raise aggregate consumer expenditure should be assessed with caution.
The results also highlight the importance of considering distributional effects (e.g., age-based response) when designing such programs.

For digital platform administration, this study provides practical insights into PLP operations implemented in digital platforms, showing how point redemption behavior affects partner firms and cross-partner transactions. Understanding these patterns helps firms develop better engagement strategies. Additionally, when public funds are distributed via PLPs, as in Japan, aligning point allocation with redemption is crucial for transparency and equitable policy outcomes.

This study has several limitations that affect the generalizability of its findings. First, the data are from users of a specific financial management application in Japan. These users may be more financially literate than the general population, and the recorded PLP data may not be complete. Japan's uniquely competitive loyalty program landscape may mean that the findings are not directly applicable to other cultural or economic contexts. Next, the quality of user-provided text data for RQ1 varied, which could affect topic extraction. The findings for RQ2 are based on a single, short-term government campaign, limiting their scope. Finally, the relationship between shopping style and point redemption explored in RQ3 is correlational, not causal. Future research should address these limitations by incorporating more diverse data sources, conducting cross-national analyses, examining long-term effects of point grants, and exploring causal links between shopping behavior and point usage. Further work could also analyze cross-redemption patterns, assess competitive dynamics among PLPs, and investigate the drivers of substitution and complementarity among partner firms using experimental or quasi-experimental methods.

\section*{Declarations}

% \subsection*{Ethics approval and consent to participate}
% Not applicable.

% \subsection*{Consent for publication}
% Not applicable.

% \subsection*{Competing interests}
% The authors declare no competing interests.

% \subsection*{Availability of data and material}
% The datasets are not publicly available due to privacy restrictions agreed upon with the data provider Kufu Company Holdings Inc. but aggregated, anonymized data may be available from the corresponding author on reasonable request and with permission of Kufu Company Holdings Inc.

\subsection*{Funding}
This research was supported by the Japan Society for the Promotion of Science (JSPS) under Grant-in-Aid for Scientific Research (JSPS KAKENHI) numbers
JP21H00756,
JP25K00680,
JP22K20159,
JP24K16359.

\subsection*{Author contributions}
A.M. conceived the study, designed the methodology, performed the analysis, and wrote the manuscript. T.T. contributed to methodology development, analysis, data acquisition, manuscript revision, and project administration. E.M. and H.T. contributed to the interpretation and manuscript revision. All authors read and approved the final manuscript.

\subsection*{Acknowledgements}
The authors are grateful for the dedicated comments and feedback from the participants at the conferences of the Japan Institute of Marketing Science.
The authors used ChatGPT to assist with English language editing and to improve the clarity of expression. The authors take full responsibility for the content of the manuscript.
\appendix
\section*{Supplementary Information}

\section{Distribution of transaction amount by demographic}\label{app:rq1_basic_stats}

\begin{figure}[!htbp]
  \centering
  \includegraphics[width=0.95\linewidth]{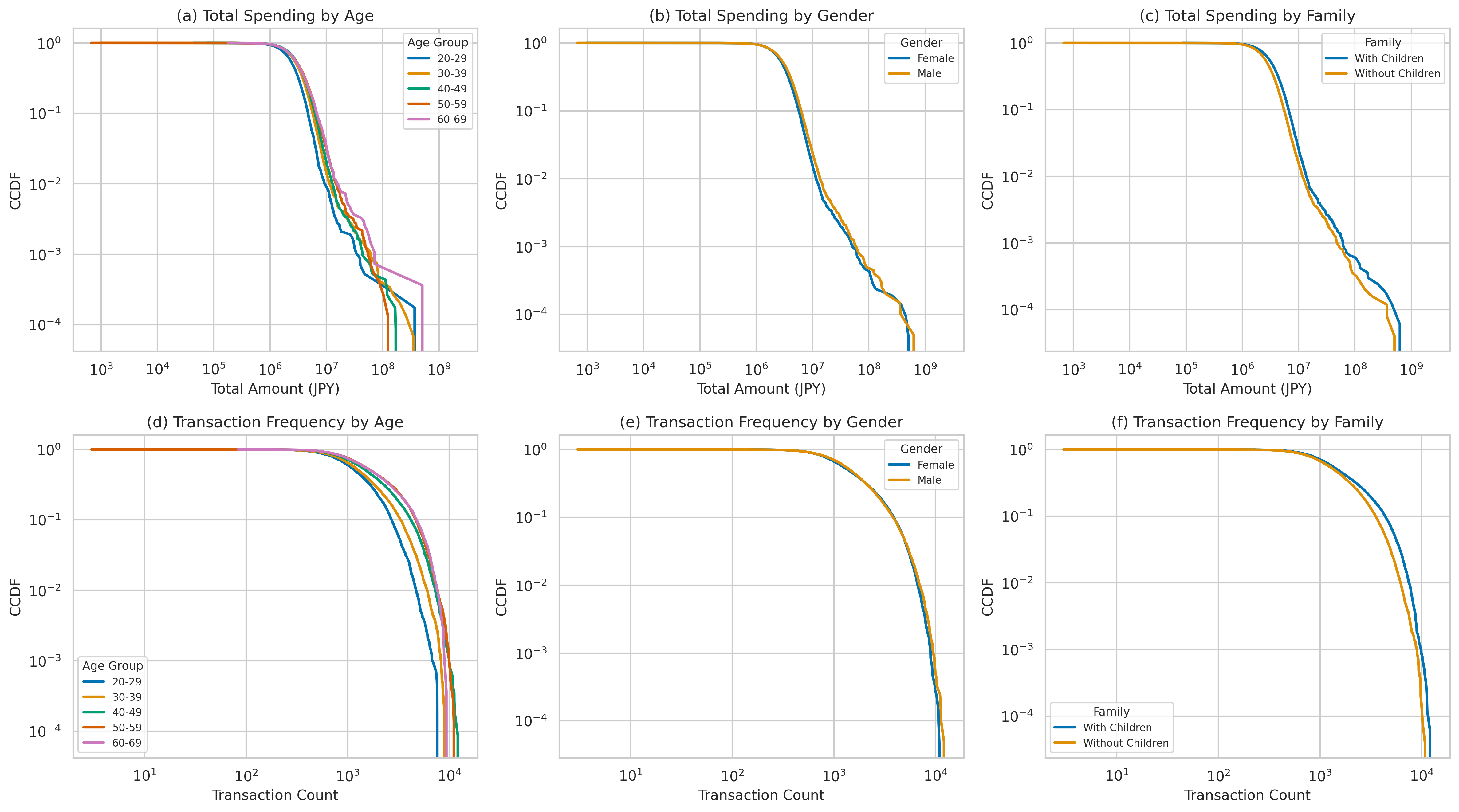}
  \caption{Distribution of transaction amount and frequency by demographic. (a-c) Complementary cumulative distribution (CCDF) of total spending amount by age group, gender, and family composition, respectively. (d-f) CCDF of transaction count by age group, gender, and family composition, respectively.}
  \label{fig:rq1_basic_stats}
\end{figure}

\section{Grant Timing Distribution}\label{app:grant_timing}

\begin{figure}[!htbp]
  \centering
  \includegraphics[width=0.6\linewidth]{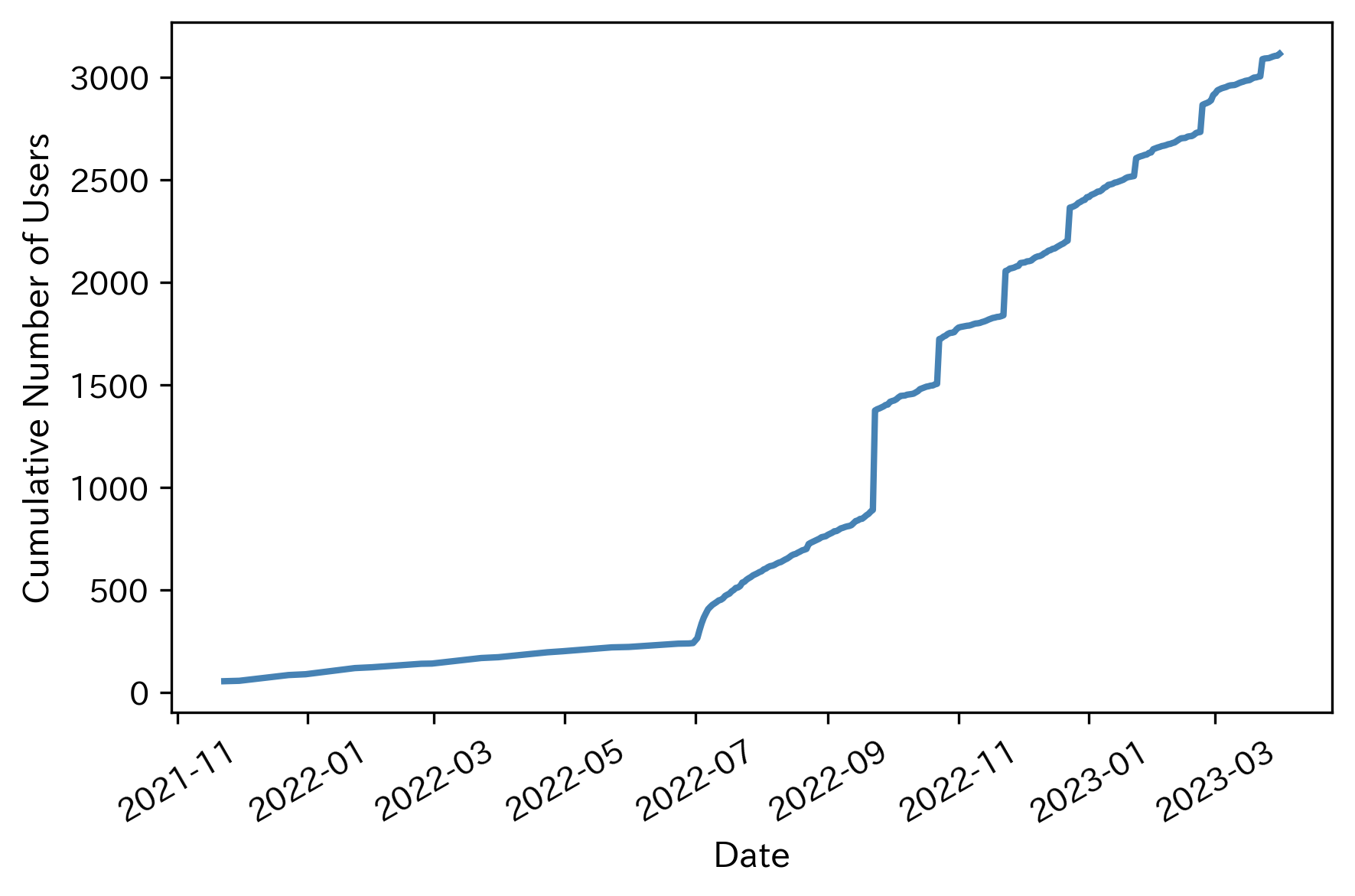}
  \caption{Cumulative number of customers who received points from the campaign, showing that 349 users received points on September 23, 2022. We treat users who received points on this date as the treatment group.}
  \label{fig:did_treatment_group}
\end{figure}

\section{Matched-Control Spending Trend}\label{app:psm_trend}

\begin{figure}[!htbp]
  \centering
  \includegraphics[width=0.7\linewidth]{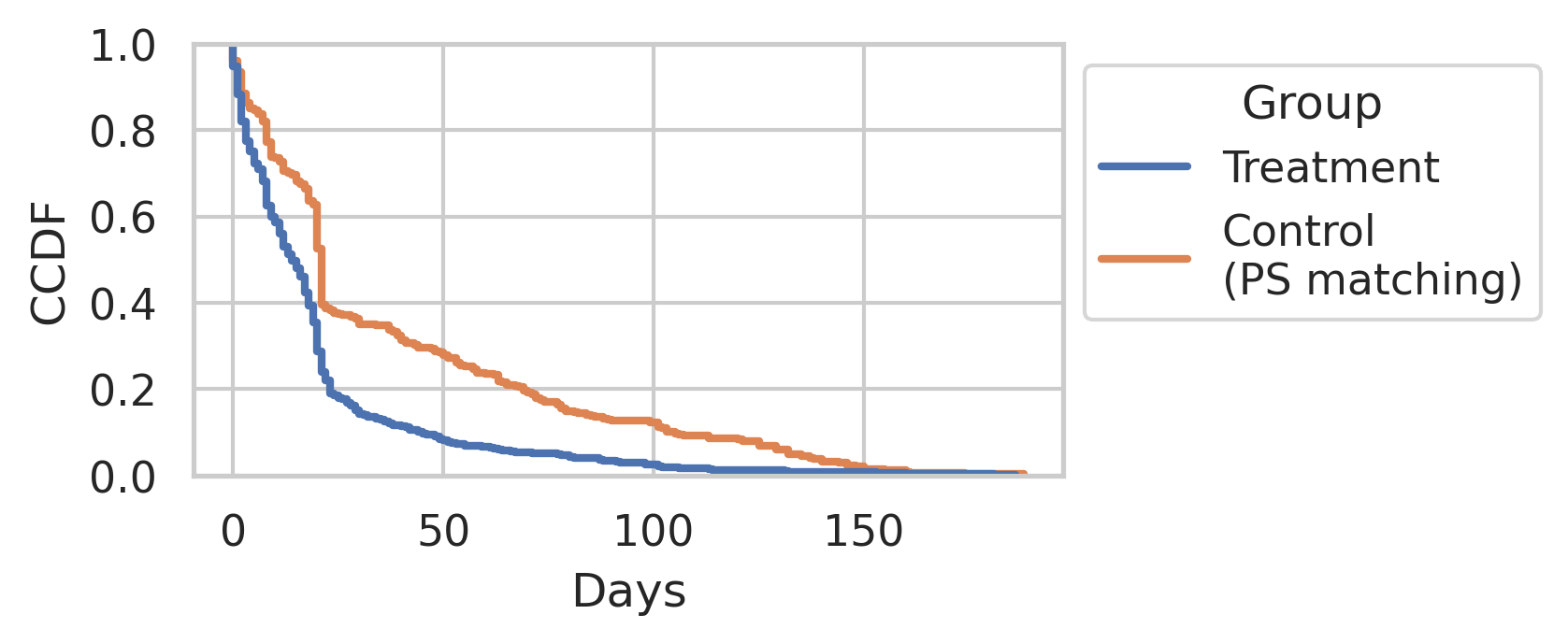}
  \caption{
    Time to cumulative 7,500-point redemption. CCDF showing the proportion of users who have not yet redeemed a cumulative 7,500 points (JPY equivalent) over days since the intervention. The control group was constructed via propensity score matching on demographics and pre-intervention spending. The leftward shift of the treatment curve indicates that the grant accelerated the pace of point redemption.
  }
  \label{fig:ps_matching_spending}
\end{figure}

\section{LDA Parameter Selection}\label{app:lda}

To select the number of topics for the LDA model, we trained models with 30, 50, and 100 topics on the preprocessed point transaction text data. For each model, we extracted the top 100 words per topic ranked by their conditional probability. These word lists were then provided to a large language model (GPT-4o) along with the contextual note that the words originated from a topic model of point transaction records; the model generated candidate labels for each topic. After manual review of the generated labels, we concluded that the 30-topic model provided the most semantically coherent and practically interpretable groupings.

Text preprocessing was performed as follows. We applied morphological analysis using GiNZA and retained only noun tokens. Place names and numerals were removed. Terms appearing in the top 0.1\% by frequency were excluded to filter out uninformative high-frequency words. The resulting token sequences were converted into a document-term matrix using scikit-learn's CountVectorizer with an upper document-frequency threshold to further remove overly common terms.

\section{Data Preprocessing Details}\label{app:preprocessing}

Store name standardization involved three steps: (1)~unification of variant spellings (e.g., different representations of the same chain store), (2)~normalization of full-width and half-width characters and whitespace, and (3)~mapping of chain-store branches to their canonical names. Category standardization was performed by mapping internal category IDs to human-readable names; transactions with missing category information were excluded from category-level analyses.

\section{Shannon Entropy Analysis}\label{app:entropy}

As a complementary measure to the embedding-based exploration scores, we computed the normalized Shannon entropy of each user's store-visit frequency distribution. For user $i$ who visited $M_i$ distinct stores with visit proportions $p_1, p_2, \ldots, p_{M_i}$, the normalized entropy is defined as:
\[
  H_i = -\frac{1}{\log M_i} \sum_{j=1}^{M_i} p_j \log p_j,
\]
where $H_i \in [0, 1]$; values near 1 indicate uniformly distributed visits (high diversity), while values near 0 indicate concentration on a few stores.

We computed this measure for the same sample used for the exploration--exploitation analysis, finding qualitatively consistent results. Pairwise Welch's $t$-tests showed that younger cohorts had significantly higher normalized entropy than older cohorts. The largest difference was between the 20--29 and 60--69 age groups ($p<0.001$), followed by differences between the 30--39 and 60--69 groups ($p<0.005$) and between the 40--49 and 60--69 groups ($p<0.02$). These results suggest that younger consumers exhibit more diverse store-visiting behavior.

\begin{figure}[!htbp]
  \centering
  \includegraphics[width=0.85\linewidth]{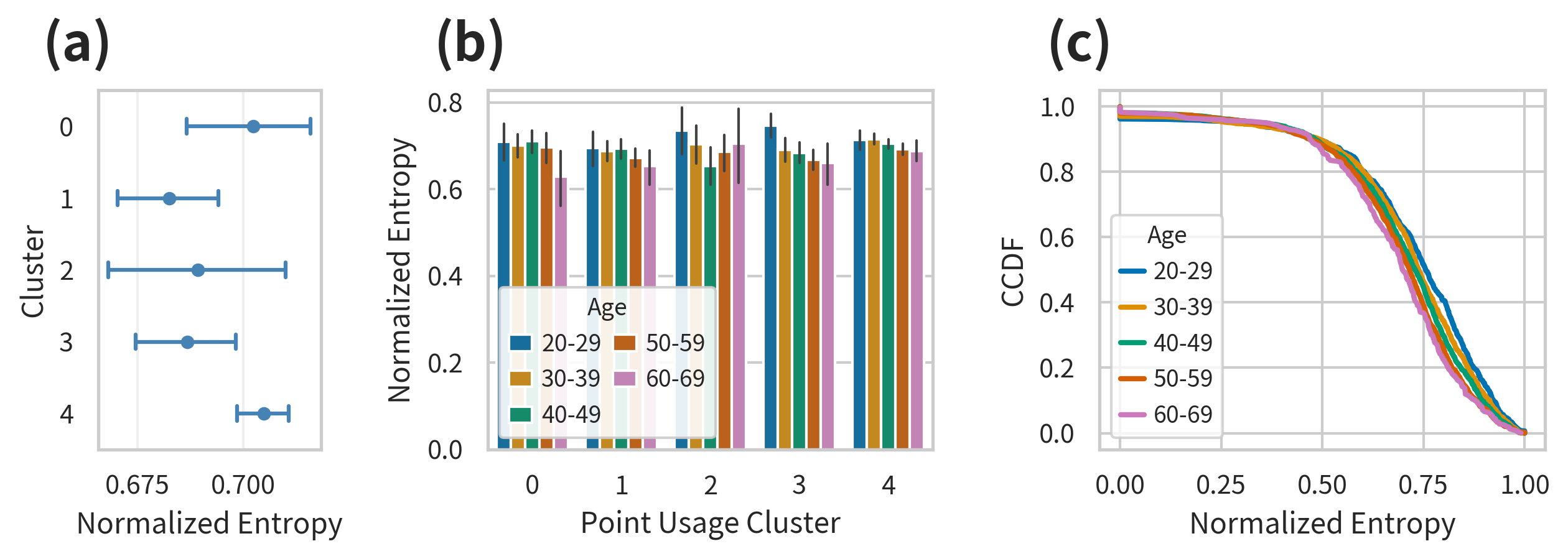}
  \caption{ Entropy analysis of shopping behavior. (a)~Average entropy across different point usage clusters; error bars represent 95\% confidence intervals. (b)~Average entropy value by point usage cluster and age group. (c)~CCDF of entropy value by user age group.}
  \label{fig:app_entropy}
\end{figure}

\section{Embedding Dimension Sensitivity Analysis}\label{app:embedding}

To assess the robustness of the exploration--exploitation scores to the embedding specification, we re-trained the SGNS (word2vec) model at three embedding dimensions: $d=64$, $d=128$ (baseline), and $d=300$. All other hyperparameters (context window $=5$, negative samples $=5$, 10 epochs) were held constant.

\input{table3}
\input{table4}

All pairwise Spearman rank correlations exceed $\rho = 0.999$, indicating that the relative ranking of users by exploration tendency is virtually identical regardless of the embedding dimension. The mean scores and standard deviations are also highly stable. These results confirm that the findings reported in the main text are not sensitive to this modeling choice.

\begin{figure}[!htbp]
  \centering
  \includegraphics[width=0.85\linewidth]{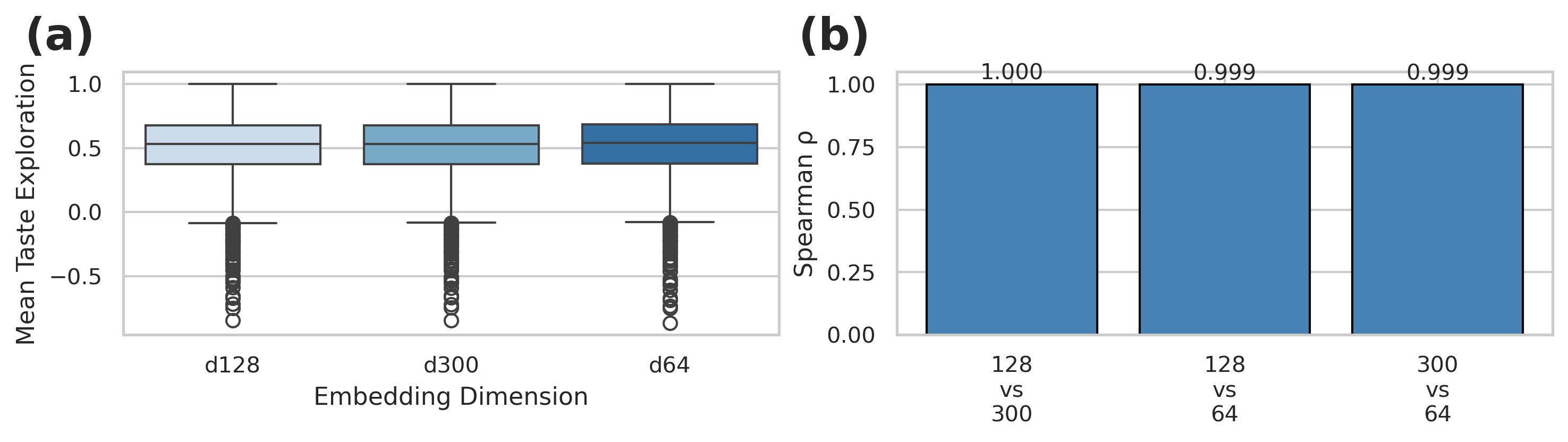}
  \caption{Embedding dimension sensitivity analysis. (a)~Distribution of exploration scores for each embedding dimension. (b)~Pairwise Spearman rank correlations plots.}
  \label{fig:app_sensitivity}
\end{figure}

\end{document}

%% file: table1.tex
\begin{table}[htbp]
    \centering
    \begin{tabular}{lp{10cm}}
    \toprule
    \textbf{Term} & \textbf{Definition} \\
    \midrule
    Partnership Loyalty Program (PLP) & Multiple firms jointly issue and redeem reward points for a shared customer base. Customers can earn and use points across participating firms. \\
    Point & A unit of value issued within a PLP, typically equivalent to 1 Japanese yen. Points are redeemable only within the same PLP. \\
    Point Earning & The act of acquiring reward points within a PLP through purchases or promotional activities. \\
    Point Redemption & The act of using accumulated reward points within a PLP to make purchases, with each point generally equivalent to 1 yen. \\
    Monetary/Cash Spending & Expenditures made using payment methods other than points (e.g., cash, credit cards). \\
    \bottomrule
    \end{tabular}
    \caption{Terminology in the Context of Partnership Loyalty Programs}
    \label{tab:plp_terms}
    \end{table}

%% file: table2.tex
\begin{table}[htbp]
    \centering
    \caption{Point usage clusters and their interpreted labels}
    \label{tab:point_cluster_labels}
    \begin{tabular}{cl}
        \toprule
        \textbf{Cluster ID} & \textbf{Interpretation} \\
        \midrule
        0 & Lifestyle-oriented mall shoppers \\
        1 & Campaign-focused commercial facility users \\
        2 & Food chain and beauty service enthusiasts \\
        3 & Dining-focused shoppers \\
        4 & Convenience store and department store loyalists \\
        \bottomrule
    \end{tabular}
\end{table}

%% file: table3.tex
\begin{table}[!htbp]
  \centering
  \caption{Summary statistics of exploration scores by embedding dimension.}
  \label{tab:app_embedding_summary}
  \begin{tabular}{lcc}
    \hline
    \textbf{Dimension} & \textbf{Mean exploration score} & \textbf{Std.\ dev.} \\
    \hline
    $d=64$  & 0.4649 & 0.2251 \\
    $d=128$ & 0.4704 & 0.2256 \\
    $d=300$ & 0.4705 & 0.2255 \\
    \hline
  \end{tabular}
\end{table}

%% file: table4.tex
\begin{table}[!htbp]
  \centering
  \caption{Pairwise Spearman rank correlations of user-level exploration scores across embedding dimensions.}
  \label{tab:app_embedding_corr}
  \begin{tabular}{lc}
    \hline
    \textbf{Pair} & \textbf{Spearman $\rho$} \\
    \hline
    $d=64$ vs.\ $d=128$  & 0.9995 \\
    $d=64$ vs.\ $d=300$  & 0.9994 \\
    $d=128$ vs.\ $d=300$ & 0.9999 \\
    \hline
  \end{tabular}
\end{table}

%% file: manuscript.bbl
\begin{thebibliography}{10}
  \urlstyle{rm}
  \expandafter\ifx\csname url\endcsname\relax
  \def\url#1{\texttt{#1}}\fi
  \expandafter\ifx\csname urlprefix\endcsname\relax\def\urlprefix{URL }\fi
  \expandafter\ifx\csname doiprefix\endcsname\relax\def\doiprefix{DOI: }\fi
  \providecommand{\bibinfo}[2]{#2}
  \providecommand{\eprint}[2][]{\url{#2}}

  \bibitem{lazer2009computational}
  \bibinfo{author}{Lazer, D.} \emph{et~al.}
  \newblock \bibinfo{journal}{\bibinfo{title}{Computational social science}}.
  \newblock {\emph{\JournalTitle{Science}}} \textbf{\bibinfo{volume}{323}}, \bibinfo{pages}{721--723} (\bibinfo{year}{2009}).

  \bibitem{blattberg2008database}
  \bibinfo{author}{Blattberg, R.~C.}, \bibinfo{author}{Kim, B.~D.} \& \bibinfo{author}{Neslin, S.~A.}
  \newblock \emph{\bibinfo{title}{Database Marketing: Analyzing and Managing Customers}} (\bibinfo{publisher}{Springer}, \bibinfo{address}{New York, NY}, \bibinfo{year}{2008}).

  \bibitem{chen2021three}
  \bibinfo{author}{Chen, Y.}, \bibinfo{author}{Mandler, T.} \& \bibinfo{author}{Meyer-Waarden, L.}
  \newblock \bibinfo{journal}{\bibinfo{title}{Three decades of research on loyalty programs: A literature review and future research agenda}}.
  \newblock {\emph{\JournalTitle{Journal of Business Research}}} \textbf{\bibinfo{volume}{124}}, \bibinfo{pages}{179--197} (\bibinfo{year}{2021}).

  \bibitem{breugelmans2015advancing}
  \bibinfo{author}{Breugelmans, E.} \emph{et~al.}
  \newblock \bibinfo{journal}{\bibinfo{title}{Advancing research on loyalty programs: a future research agenda}}.
  \newblock {\emph{\JournalTitle{Marketing Letters}}} \textbf{\bibinfo{volume}{26}}, \bibinfo{pages}{127--139} (\bibinfo{year}{2015}).

  \bibitem{dorotic2021synergistic}
  \bibinfo{author}{Dorotic, M.}, \bibinfo{author}{Fok, D.}, \bibinfo{author}{Verhoef, P.~C.} \& \bibinfo{author}{Bijmolt, T. H.~A.}
  \newblock \bibinfo{journal}{\bibinfo{title}{Synergistic and cannibalization effects in a partnership loyalty program}}.
  \newblock {\emph{\JournalTitle{Journal of the Academy of Marketing Science}}} \textbf{\bibinfo{volume}{49}}, \bibinfo{pages}{1021--1042} (\bibinfo{year}{2021}).

  \bibitem{bombaij2020when}
  \bibinfo{author}{Bombaij, N. J.~F.} \& \bibinfo{author}{Dekimpe, M.~G.}
  \newblock \bibinfo{journal}{\bibinfo{title}{When do loyalty programs work? the moderating role of design, retailer--strategy, and country characteristics}}.
  \newblock {\emph{\JournalTitle{International Journal of Research in Marketing}}} \textbf{\bibinfo{volume}{37}}, \bibinfo{pages}{175--195} (\bibinfo{year}{2020}).

  \bibitem{lemon2009reinforcing}
  \bibinfo{author}{Lemon, K.~N.} \& \bibinfo{author}{Wangenheim, F.~V.}
  \newblock \bibinfo{journal}{\bibinfo{title}{The reinforcing effects of loyalty program partnerships and core service usage: a longitudinal analysis}}.
  \newblock {\emph{\JournalTitle{Journal of Service Research}}} \textbf{\bibinfo{volume}{11}}, \bibinfo{pages}{357--370} (\bibinfo{year}{2009}).

  \bibitem{kivetz2006goal}
  \bibinfo{author}{Kivetz, R.}, \bibinfo{author}{Urminsky, O.} \& \bibinfo{author}{Zheng, Y.}
  \newblock \bibinfo{journal}{\bibinfo{title}{The goal-gradient hypothesis resurrected: Purchase acceleration, illusionary goal progress, and customer retention}}.
  \newblock {\emph{\JournalTitle{Journal of Marketing Research}}} \textbf{\bibinfo{volume}{43}}, \bibinfo{pages}{39--58} (\bibinfo{year}{2006}).

  \bibitem{nunes2006endowed}
  \bibinfo{author}{Nunes, J.~C.} \& \bibinfo{author}{Dr{\'e}ze, X.}
  \newblock \bibinfo{journal}{\bibinfo{title}{The endowed progress effect: How artificial advancement increases effort}}.
  \newblock {\emph{\JournalTitle{Journal of Consumer Research}}} \textbf{\bibinfo{volume}{32}}, \bibinfo{pages}{504--512} (\bibinfo{year}{2006}).

  \bibitem{kivetz2002earning}
  \bibinfo{author}{Kivetz, R.} \& \bibinfo{author}{Simonson, I.}
  \newblock \bibinfo{journal}{\bibinfo{title}{Earning the right to indulge: Effort as a determinant of customer preferences toward frequency program rewards}}.
  \newblock {\emph{\JournalTitle{Journal of Marketing Research}}} \textbf{\bibinfo{volume}{39}}, \bibinfo{pages}{155--170} (\bibinfo{year}{2002}).

  \bibitem{kim2021emerging}
  \bibinfo{author}{Kim, J.~J.}, \bibinfo{author}{Steinhoff, L.} \& \bibinfo{author}{Palmatier, R.~W.}
  \newblock \bibinfo{journal}{\bibinfo{title}{An emerging theory of loyalty program dynamics}}.
  \newblock {\emph{\JournalTitle{Journal of the Academy of Marketing Science}}} \textbf{\bibinfo{volume}{49}}, \bibinfo{pages}{71--95} (\bibinfo{year}{2021}).

  \bibitem{jin2014when}
  \bibinfo{author}{Jin, L.} \& \bibinfo{author}{Huang, Y.}
  \newblock \bibinfo{journal}{\bibinfo{title}{When giving money does not work: The differential effects of monetary versus in-kind rewards in referral reward programs}}.
  \newblock {\emph{\JournalTitle{International Journal of Research in Marketing}}} \textbf{\bibinfo{volume}{31}}, \bibinfo{pages}{107--116} (\bibinfo{year}{2014}).


\bibitem{MizuhoBank2025PointEconomy}
\bibinfo{author}{{Mizuho Bank, Industry Research Department}}.
\newblock \bibinfo{title}{{Ts\=ushin sangy\=o no tenb\=o 2025 [Outlook for the telecommunications industry 2025]}}.
\newblock \bibinfo{type}{{Mizuho Industry Research}}~\bibinfo{number}{77}, \bibinfo{institution}{{Mizuho Bank, Ltd.}} (\bibinfo{year}{2025}).
\newblock \bibinfo{note}{In Japanese}.
\newblock \urlprefix\url{https://www.mizuhobank.co.jp/corporate/industry/sangyou/m1077.html}. Accessed 20 Jun 2026.

\bibitem{steinberg2025incentives}
\bibinfo{author}{Steinberg, M.}
\newblock \bibinfo{journal}{\bibinfo{title}{The incentives of paypay against the convenience of cash: On the conveniencing of cashless payments in japan}}.
\newblock {\emph{\JournalTitle{Platforms \& Society}}} \textbf{\bibinfo{volume}{2}}, \bibinfo{pages}{29768624251355164} (\bibinfo{year}{2025}).

\bibitem{Tomita2023LoyaltyPoints}
\bibinfo{author}{Tomita, K.}
\newblock \bibinfo{title}{{Four determinants of loyalty points' effectiveness}}.
\newblock \bibinfo{journal}{lakyara} \bibinfo{volume}{369},
\bibinfo{institution}{Nomura Research Institute, Ltd.}
\newblock \bibinfo{howpublished}{\url{https://www.nri.com/content/900013720.pdf}} (\bibinfo{year}{2023}).
\newblock \bibinfo{note}{Accessed Mar. 6, 2026.}

\bibitem{digitalAgency2022MynaPoints}
\bibinfo{author}{{Digital Agency and Ministry of Internal Affairs and Communications}}.
\newblock \bibinfo{title}{{Receive Myna Points by obtaining a My Number Card and registering a public money receiving account}}.
\newblock \bibinfo{howpublished}{\url{https://www.digital.go.jp/assets/contents/node/basic_page/field_ref_resources/73b1f4d8-b963-416a-a2e7-ef6556c2aa89/8e268260/20220525_policies_account_registration_outline_01.pdf}} (\bibinfo{year}{2022}).
\newblock \bibinfo{note}{Published May 25, 2022. Accessed Mar. 6, 2026. In Japanese.}

\bibitem{fujiki2025cashless}
\bibinfo{author}{Fujiki, H.}
\newblock \bibinfo{journal}{\bibinfo{title}{Cashless payment methods and covid-19: evidence from japanese consumer panel data}}.
\newblock {\emph{\JournalTitle{The Japanese Economic Review}}} \textbf{\bibinfo{volume}{76}}, \bibinfo{pages}{121--162} (\bibinfo{year}{2025}).

\bibitem{sekine2022going}
\bibinfo{author}{Sekine, T.}, \bibinfo{author}{Shoji, T.} \& \bibinfo{author}{Watanabe, T.}
\newblock \bibinfo{journal}{\bibinfo{title}{Going cashless: Government’s point reward program vs. covid-19}}.
\newblock {\emph{\JournalTitle{Research Project on Central Bank Communication Working Paper Series}}}  (\bibinfo{year}{2022}).
  \bibitem{sharp1997loyalty}
  \bibinfo{author}{Sharp, B.} \& \bibinfo{author}{Sharp, A.}
  \newblock \bibinfo{journal}{\bibinfo{title}{Loyalty programs and their impact on repeat-purchase loyalty patterns}}.
  \newblock {\emph{\JournalTitle{International Journal of Research in Marketing}}} \textbf{\bibinfo{volume}{14}}, \bibinfo{pages}{473--486} (\bibinfo{year}{1997}).

  \bibitem{dorotic2011do}
  \bibinfo{author}{Dorotic, M.}, \bibinfo{author}{Fok, D.}, \bibinfo{author}{Verhoef, P.~C.} \& \bibinfo{author}{Bijmolt, T. H.~A.}
  \newblock \bibinfo{journal}{\bibinfo{title}{Do vendors benefit from promotions in a multi--vendor loyalty program?}}
  \newblock {\emph{\JournalTitle{Marketing Letters}}} \textbf{\bibinfo{volume}{22}}, \bibinfo{pages}{341--356} (\bibinfo{year}{2011}).

  \bibitem{kim2015effects}
  \bibinfo{author}{Kim, S.~J.}, \bibinfo{author}{Wang, R. J.~H.} \& \bibinfo{author}{Malthouse, E.~C.}
  \newblock \bibinfo{journal}{\bibinfo{title}{The effects of adopting and using a brand's mobile application on customers' subsequent purchase behavior}}.
  \newblock {\emph{\JournalTitle{Journal of Interactive Marketing}}} \textbf{\bibinfo{volume}{31}}, \bibinfo{pages}{28--41} (\bibinfo{year}{2015}).

  \bibitem{wang2018when}
  \bibinfo{author}{Wang, R. J.~H.}, \bibinfo{author}{Krishnamurthi, L.} \& \bibinfo{author}{Malthouse, E.~C.}
  \newblock \bibinfo{journal}{\bibinfo{title}{When reward convenience meets a mobile app: Increasing customer participation in a coalition loyalty program}}.
  \newblock {\emph{\JournalTitle{Journal of the Association for Consumer Research}}} \textbf{\bibinfo{volume}{3}}, \bibinfo{pages}{314--329} (\bibinfo{year}{2018}).

  \bibitem{villacemolinero2016multi}
  \bibinfo{author}{Villac{\'e}-Molinero, T.}, \bibinfo{author}{Reinares-Lara, P.} \& \bibinfo{author}{Reinares-Lara, E.}
  \newblock \bibinfo{journal}{\bibinfo{title}{Multi-vendor loyalty programs: influencing customer behavioral loyalty?}}
  \newblock {\emph{\JournalTitle{Frontiers in Psychology}}} \textbf{\bibinfo{volume}{7}}, \bibinfo{pages}{204} (\bibinfo{year}{2016}).

  \bibitem{keh2006do}
  \bibinfo{author}{Keh, H.~T.} \& \bibinfo{author}{Lee, Y.~H.}
  \newblock \bibinfo{journal}{\bibinfo{title}{Do reward programs build loyalty for services? the moderating effect of satisfaction on type and timing of rewards}}.
  \newblock {\emph{\JournalTitle{Journal of Retailing}}} \textbf{\bibinfo{volume}{82}}, \bibinfo{pages}{127--136} (\bibinfo{year}{2006}).

  \bibitem{yi2003effects}
  \bibinfo{author}{Yi, Y.} \& \bibinfo{author}{Jeon, H.}
  \newblock \bibinfo{journal}{\bibinfo{title}{Effects of loyalty programs on value perception, program loyalty, and brand loyalty}}.
  \newblock {\emph{\JournalTitle{Journal of the Academy of Marketing Science}}} \textbf{\bibinfo{volume}{31}}, \bibinfo{pages}{229--240} (\bibinfo{year}{2003}).

  \bibitem{yao2012determining}
  \bibinfo{author}{Yao, S.}, \bibinfo{author}{Mela, C.~F.}, \bibinfo{author}{Chiang, J.} \& \bibinfo{author}{Chen, Y.}
  \newblock \bibinfo{journal}{\bibinfo{title}{Determining consumers' discount rates with field studies}}.
  \newblock {\emph{\JournalTitle{Journal of Marketing Research}}} \textbf{\bibinfo{volume}{49}}, \bibinfo{pages}{822--841} (\bibinfo{year}{2012}).

  \bibitem{thaler1985mental}
  \bibinfo{author}{Thaler, R.}
  \newblock \bibinfo{journal}{\bibinfo{title}{Mental accounting and consumer choice}}.
  \newblock {\emph{\JournalTitle{Marketing science}}} \textbf{\bibinfo{volume}{4}}, \bibinfo{pages}{199--214} (\bibinfo{year}{1985}).

  \bibitem{kivetz2003idiosyncratic}
  \bibinfo{author}{Kivetz, R.} \& \bibinfo{author}{Simonson, I.}
  \newblock \bibinfo{journal}{\bibinfo{title}{The idiosyncratic fit heuristic: Effort advantage as a determinant of consumer response to loyalty programs}}.
  \newblock {\emph{\JournalTitle{Journal of Marketing Research}}} \textbf{\bibinfo{volume}{40}}, \bibinfo{pages}{454--467} (\bibinfo{year}{2003}).

  \bibitem{dreze1998exploiting}
  \bibinfo{author}{Dr{\'e}ze, X.} \& \bibinfo{author}{Hoch, S.~J.}
  \newblock \bibinfo{journal}{\bibinfo{title}{Exploiting the installed base using cross-merchandising and category destination programs}}.
  \newblock {\emph{\JournalTitle{International Journal of Research in Marketing}}} \textbf{\bibinfo{volume}{15}}, \bibinfo{pages}{459--471} (\bibinfo{year}{1998}).

  \bibitem{dreze2004using}
  \bibinfo{author}{Dr{\'e}ze, X.} \& \bibinfo{author}{Nunes, J.~C.}
  \newblock \bibinfo{journal}{\bibinfo{title}{Using combined-currency prices to lower consumers' perceived cost}}.
  \newblock {\emph{\JournalTitle{Journal of Marketing Research}}} \textbf{\bibinfo{volume}{41}}, \bibinfo{pages}{59--72} (\bibinfo{year}{2004}).

  \bibitem{dreze2011recurring}
  \bibinfo{author}{Dr{\'e}ze, X.} \& \bibinfo{author}{Nunes, J.~C.}
  \newblock \bibinfo{journal}{\bibinfo{title}{Recurring goals and learning: The impact of successful reward attainment on purchase behavior}}.
  \newblock {\emph{\JournalTitle{Journal of Marketing Research}}} \textbf{\bibinfo{volume}{48}}, \bibinfo{pages}{268--281} (\bibinfo{year}{2011}).

  \bibitem{stourm2015stockpiling}
  \bibinfo{author}{Stourm, V.}, \bibinfo{author}{Bradlow, E.~T.} \& \bibinfo{author}{Fader, P.~S.}
  \newblock \bibinfo{journal}{\bibinfo{title}{Stockpiling points in linear loyalty programs}}.
  \newblock {\emph{\JournalTitle{Journal of Marketing Research}}} \textbf{\bibinfo{volume}{52}}, \bibinfo{pages}{253--267} (\bibinfo{year}{2015}).

  \bibitem{noble2014accumulation}
  \bibinfo{author}{Noble, S.~M.}, \bibinfo{author}{Esmark, C.~L.} \& \bibinfo{author}{Noble, C.~H.}
  \newblock \bibinfo{journal}{\bibinfo{title}{Accumulation versus instant loyalty programs: The influence of controlling policies on customers' commitments}}.
  \newblock {\emph{\JournalTitle{Journal of Business Research}}} \textbf{\bibinfo{volume}{67}}, \bibinfo{pages}{361--368} (\bibinfo{year}{2014}).

  \bibitem{kivetz2005promotion}
  \bibinfo{author}{Kivetz, R.}
  \newblock \bibinfo{journal}{\bibinfo{title}{Promotion reactance: The role of effort‐reward congruity}}.
  \newblock {\emph{\JournalTitle{Journal of Consumer Research}}} \textbf{\bibinfo{volume}{31}}, \bibinfo{pages}{725--736} (\bibinfo{year}{2005}).

  \bibitem{Blei2003}
  \bibinfo{author}{Blei, D.~M.}, \bibinfo{author}{Ng, A.~Y.} \& \bibinfo{author}{Jordan, M.~I.}
  \newblock \bibinfo{journal}{\bibinfo{title}{Latent dirichlet allocation}}.
  \newblock {\emph{\JournalTitle{Journal of machine Learning research}}} \textbf{\bibinfo{volume}{3}}, \bibinfo{pages}{993--1022} (\bibinfo{year}{2003}).

  \bibitem{SantAnnaZhao2020}
  \bibinfo{author}{Sant’Anna, P.~H.} \& \bibinfo{author}{Zhao, J.}
  \newblock \bibinfo{journal}{\bibinfo{title}{Doubly robust difference-in-differences estimators}}.
  \newblock {\emph{\JournalTitle{Journal of econometrics}}} \textbf{\bibinfo{volume}{219}}, \bibinfo{pages}{101--122} (\bibinfo{year}{2020}).

  \bibitem{Kim2024PNAS}
  \bibinfo{author}{Kim, K.}, \bibinfo{author}{Askin, N.} \& \bibinfo{author}{Evans, J.~A.}
  \newblock \bibinfo{journal}{\bibinfo{title}{Disrupted routines anticipate musical exploration}}.
  \newblock {\emph{\JournalTitle{Proceedings of the National Academy of Sciences}}} \textbf{\bibinfo{volume}{121}}, \bibinfo{pages}{e2306549121} (\bibinfo{year}{2024}).

\bibitem{GOMEZZARA2024108014}
\bibinfo{author}{G{\'o}mez-Zar{\'a}, D.}, \bibinfo{author}{Liu, Y.}, \bibinfo{author}{Neves, L.}, \bibinfo{author}{Shah, N.} \& \bibinfo{author}{Bos, M.~W.}
\newblock \bibinfo{journal}{\bibinfo{title}{Unpacking the exploration--exploitation tradeoff on snapchat: The relationships between users' exploration--exploitation interests and server log data}}.
\newblock {\emph{\JournalTitle{Computers in Human Behavior}}} \textbf{\bibinfo{volume}{150}}, \bibinfo{pages}{108014} (\bibinfo{year}{2024}).


  \bibitem{mok2022dynamics}
  \bibinfo{author}{Mok, L.}, \bibinfo{author}{Way, S.~F.}, \bibinfo{author}{Maystre, L.} \& \bibinfo{author}{Anderson, A.}
  \newblock \bibinfo{title}{The dynamics of exploration on spotify}.
  \newblock In \emph{\bibinfo{booktitle}{Proceedings of the International AAAI Conference on Web and Social Media}}, vol.~\bibinfo{volume}{16}, \bibinfo{pages}{663--674} (\bibinfo{year}{2022}).

  \bibitem{waller2021quantifying}
  \bibinfo{author}{Waller, I.} \& \bibinfo{author}{Anderson, A.}
  \newblock \bibinfo{journal}{\bibinfo{title}{Quantifying social organization and political polarization in online platforms}}.
  \newblock {\emph{\JournalTitle{Nature}}} \textbf{\bibinfo{volume}{600}}, \bibinfo{pages}{264--268} (\bibinfo{year}{2021}).

  \bibitem{mikolov2013efficient}
  \bibinfo{author}{Mikolov, T.}, \bibinfo{author}{Chen, K.}, \bibinfo{author}{Corrado, G.} \& \bibinfo{author}{Dean, J.}
  \newblock \bibinfo{title}{Efficient estimation of word representations in vector space}.
  \newblock In \emph{\bibinfo{booktitle}{International Conference on Learning Representations}} (\bibinfo{year}{2013}).

  \bibitem{mikolov2013dist}
  \bibinfo{author}{Mikolov, T.}, \bibinfo{author}{Sutskever, I.}, \bibinfo{author}{Chen, K.}, \bibinfo{author}{Corrado, G.} \& \bibinfo{author}{Dean, J.}
  \newblock \bibinfo{title}{Distributed representations of words and phrases and their compositionality}.
  \newblock In \emph{\bibinfo{booktitle}{Advances in Neural Information Processing Systems}} (\bibinfo{year}{2013}).

  \bibitem{levy2014dependency}
  \bibinfo{author}{Levy, O.} \& \bibinfo{author}{Goldberg, Y.}
  \newblock \bibinfo{title}{Dependency-based word embeddings}.
  \newblock In \emph{\bibinfo{booktitle}{Proceedings of the 52nd Annual Meeting of the Association for Computational Linguistics}} (\bibinfo{year}{2014}).

  \bibitem{ChenPointRedemption2024}
  \bibinfo{author}{Li, C.}, \bibinfo{author}{Swaminathan, S.} \& \bibinfo{author}{Kim, J.}
  \newblock \bibinfo{journal}{\bibinfo{title}{Point redemption in loyalty programs: the role of customer relationship characteristics and their implications for service providers}}.
  \newblock {\emph{\JournalTitle{Journal of Service Research}}}  (\bibinfo{year}{2024}).

\end{thebibliography}
